\documentclass[aps, prl, superscriptaddress, reprint, twocolumn, amsmath, amssymb, notitlepage]{revtex4-2}
\usepackage{graphicx}
\usepackage{amsmath,amssymb}
\usepackage{color}
\usepackage{mathrsfs}
\usepackage{physics}
\usepackage[utf8x]{inputenc}

\newcommand{\ptk}[1]{#1}
\newcommand{\comment}[1]{{}}
\newcommand{\rev}[1]{#1}

\begin{document}
\title{
Cavity-induced exciton localisation and photon blockade in two-dimensional semiconductors coupled to an electromagnetic resonator
}

\author{Emil V. Denning}
\email{emvo@fotonik.dtu.dk}
\affiliation{Department of Photonics Engineering, Technical University of Denmark, 2800 Kgs. Lyngby, Denmark}
\affiliation{NanoPhoton - Center for Nanophotonics, Technical University of Denmark, Ørsteds Plads 345A, DK-2800 Kgs. Lyngby, Denmark}
\affiliation{Nichtlineare Optik und Quantenelektronik, Institut f\"ur Theoretische Physik, Technische Universit\"at Berlin, 10623 Berlin, Germany}

\author{Martijn Wubs}
\affiliation{Department of Photonics Engineering, Technical University of Denmark, 2800 Kgs. Lyngby, Denmark}
\affiliation{NanoPhoton - Center for Nanophotonics, Technical University of Denmark, Ørsteds Plads 345A, DK-2800 Kgs. Lyngby, Denmark}
\affiliation{Center for Nanostructured Graphene, Technical University of Denmark, 2800 Kgs. Lyngby, Denmark}

\author{Nicolas Stenger}
\affiliation{Department of Photonics Engineering, Technical University of Denmark, 2800 Kgs. Lyngby, Denmark}
\affiliation{NanoPhoton - Center for Nanophotonics, Technical University of Denmark, Ørsteds Plads 345A, DK-2800 Kgs. Lyngby, Denmark}
\affiliation{Center for Nanostructured Graphene, Technical University of Denmark, 2800 Kgs. Lyngby, Denmark}

\author{Jesper M\o rk}
\affiliation{Department of Photonics Engineering, Technical University of Denmark, 2800 Kgs. Lyngby, Denmark}
\affiliation{NanoPhoton - Center for Nanophotonics, Technical University of Denmark, Ørsteds Plads 345A, DK-2800 Kgs. Lyngby, Denmark}

\author{Philip Trøst Kristensen}
\affiliation{Department of Photonics Engineering, Technical University of Denmark, 2800 Kgs. Lyngby, Denmark}
\affiliation{NanoPhoton - Center for Nanophotonics, Technical University of Denmark, Ørsteds Plads 345A, DK-2800 Kgs. Lyngby, Denmark}

\date{\today}

\begin{abstract}
Recent experiments have demonstrated strong light--matter coupling between \ptk{electromagnetic} %optical
nanoresonators and \ptk{pristine %monolayer
sheets} of \ptk{two-dimensional} %(2D)
semiconductors, and it has been speculated whether these systems can enter the quantum regime operating at the few-polariton level. \ptk{To address this question,} %Here
we present a microscopic quantum theory for the interaction between excitons in %a translation-invariant
\ptk{a sheet of two-dimensional material} %2D sheet
and a localised electromagnetic resonator. We find that the light-matter interaction breaks the symmetry of the otherwise translation-invariant system and thereby effectively generates a localised exciton mode, which is coupled to an environment of residual exciton modes. This dissiepative coupling increases with tighter lateral confinement, and our analysis reveals this to be a potential challenge in realising nonlinear exciton-exciton interaction.
Nonetheless, we predict that polariton blockade due to nonlinear exciton-exciton interactions is \ptk{well} within reach for \ptk{nanoresonators} %cavities
coupled to transition-metal dichalcogenides, provided that the lateral \rev{confinement} %\rev{electromagnetic} %resonator mode confinement
can be sufficiently \rev{tight to make} %small that
the nonlinearity overcome the polariton dephasing caused by phonon interactions.
\end{abstract}
	\maketitle

%\paragraph{Introduction \textemdash}

Interfacing an electromagnetic resonator with the spatially extended excitons in a pristine sheet of two-dimensional (2D) semiconductor has recently been demonstrated as a way of obtaining very large Rabi splittings, up to the order of 100 meV~\cite{wen2017room,zheng2017manipulating,kleemann2017strong,cuadra2018observation,stuhrenberg2018strong,han2018rabi,geisler2019single,qin2020revealing}. These strong interactions are of great interest due to the prospect of realising polaritonic devices~\cite{sanvitto2016road} such as squeezed-light sources~\cite{karr2004squeezing,boulier2014polariton}, polariton lasers~\cite{imamog1996nonequilibrium} and polariton \rev{blockade} which\ptk{, in turn, enables the construction of single-photon sources}% allows for
%single-photon generation%
~\cite{verger2006polariton,ferretti2012single,delteil2019towards,munoz2019emergence} and few-photon logic gates~\cite{chang2007single,volz2012ultrafast,hwang2009single}. Despite a growing interest in these systems, \ptk{there is a lack of microscopic modeling of the experiments, which} are colloquially analysed \ptk{by use of} phenomenological coupled-oscillator models\ptk{. Even if these models can be well fitted to experimental data, the lack of a microscopic foundation limits their ability to predict quantum optical figures of merit of practical interest, such as second-order correlation functions.}

%, which offer few predictive insights to help in designing suitable

%offering few predictive insights.} % to the polaritonic nature
%with no apparent microscopic justification and offering little}. %which appear to be unsuited for

\begin{figure}[tb]
  \centering
  \includegraphics[width=\columnwidth]{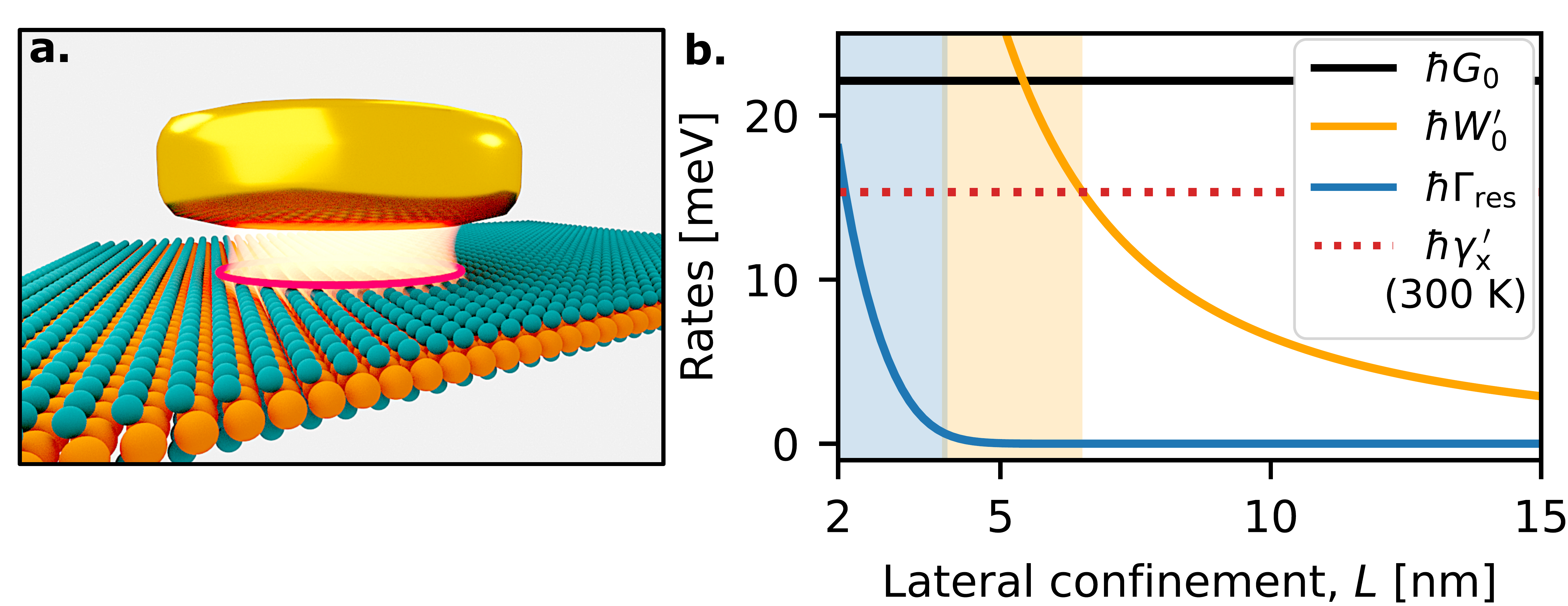}
  \caption{{\bf a.} Illustration of \ptk{an electromagnetic} %optical
    resonator coupled to a sheet of 2D material. {\bf b.} Exciton--resonator coupling $G_0$, nonlinear exciton-exciton interaction $W_0'$, and decay rate into residual exciton modes, $\Gamma_{\rm res}$, as a function of the lateral confinement, $L$, of the electromagnetic field, as well as the exciton dephasing from phonon scattering, $\gamma_{\rm x}^{\:\prime}$, in $\mathrm{WS_2}$ at a temperature of 300 K. Yellow shading marks the regime where $L$ is large enough that the residual excitons can be ignored and small enough that $W_0'$ exceeds the polariton dephasing, such that polariton blockade is possible. Blue shading marks the regime where the residual excitons have a non-negligible impact on the dynamics.}
  \label{fig:Fig_1_frontpage}
%\vspace{-5mm}
\end{figure}

%\ptk{To address the question of whether systems of 2D semiconductors coupled to electromagnetic restonators can operate as polaritonic devices, in this Article we} %
In this \ptk{Letter}, we %paper, we
develop a microscopic quantum theory \ptk{of 2D semiconductors coupled to electromagnetic resonators}, which consistently links important dynamical quantities --- such as the exciton--resonator coupling strength and effective nonlinear exciton-exciton interaction --- to fundamental material parameters. The theory applies equally to plasmonic resonators~\cite{wen2017room,zheng2017manipulating,kleemann2017strong,cuadra2018observation,stuhrenberg2018strong,han2018rabi,geisler2019single,qin2020revealing} and dielectric nanocavities~\cite{wu2014control,noori2016photonic,fryett2016silicon}, including a new generation of dielectric cavities with extreme confinement of light~\cite{hu2016design,choi2017self,wang2018maximizing}.
Our approach is based on 2D Wannier-Mott excitons with discrete translation invariance broken only by interactions with \ptk{the} localised \ptk{electromagnetic resonator, as illustrated in Fig.~\ref{fig:Fig_1_frontpage}a.} %mode
%(see Fig.~\ref{fig:intro}a-b).
We find that the symmetry-breaking light--matter interaction generates a collective localised exciton mode %(see Fig.~\ref{fig:intro}c)
-- an \emph{exciton reaction coordinate} -- defined by the \ptk{profile of the resonant electric field}. \ptk{The exciton reaction coordinate, in turn,} is coupled to \ptk{an} environment of residual exciton modes with a spectral density that depends on the lateral extent of the \rev{resonant} %electromagnetic
field. Coupling to the residual exciton \ptk{modes is pronounced only for very tight lateral confinement of the electric field} %, and that it}
and can be neglected for %sufficiently large lateral optical
confinement lengths above a few nanometers \ptk{in} %for
realistic systems\ptk{, as illustrated Fig.~\ref{fig:Fig_1_frontpage}b.} % (see Fig.~\ref{fig:intro}d). %On the other hand,
\ptk{The} effective nonlinear exciton-exciton interaction strength in the \ptk{reaction coordinate} %collective exciton mode
also increases with decreasing lateral mode area, % of the optical mode,
thereby making structures with tight lateral confinement crucial for
realizing polaritonic devices. Intriguingly, we
find that there exists a regime where the lateral confinement %length scale
is \ptk{tight} %small
enough to %generate a significant nonlinear response that
\ptk{enable} polariton blockade \ptk{operation}, yet sufficiently large to avoid coupling to residual excitons. In this regime, therefore, one can unambiguously interpret the dynamics in terms of nonlinear multi-exciton interactions within the single reaction coordinate. As a main result, we %We
predict that polariton blockade can be reached using a monolayer transition-metal dichalcogenide coupled to an electromagnetic resonator, since the nonlinearity can exceed the dephasing caused by thermal phonon interactions. %\textcolor{red}{The theory goes beyond the dipole approximation, and we derive the important result that the coupling strength is independent of the lateral mode confinement, $L$, and only depends on the optical confinement in the out-of-plane direction (see Fig.~\ref{fig:intro}d). This finding is in contrast with conventional wisdom, stating that the coupling strength scales as $1/\sqrt{V}$, where $V$ is the optical mode volume: This scaling is only valid within the dipole approximation, which often applies for quantum emitters with highly localised excitons, such as quantum dots~\cite{lodahl2015interfacing} or crystal defects~\cite{jelezko2006single}, but does not apply to the case of pristine 2D-materials, where the exciton eigenstates are spatially delocalised momentum states. The independence of lateral confinement stems from the fact that the spatial distribution of the exciton reaction coordinate is defined by the optical mode, which leads to perfect lateral co-localisation of excitons and photons. Furthermore, the theory gives access to the nonlinear exciton-exciton interaction strength in the collective exciton mode, which scales as $1/L^2$ (see Fig.~\ref{fig:intro}d). We assess the prospects of reaching polariton blockade in nanocavities coupled to a pristine transition-metal dichalcogenide sheet and show that this is feasible when the nonlinearity overcomes the excitonic pure dephasing caused by thermal phonon interactions.}
In contrast to earlier work in the context of quantum wells~\cite{verger2006polariton} and 2D-materials~\cite{tserkezis2020applicability}, the \rev{present microscopic model is based on}  %is constructed microscopically using
extended exciton momentum-states, thereby connecting all model parameters directly to measurable material properties and identifying the regime where the influence of the residual excitons cannot be neglected. \rev{The model accounts for general electromagnetic resonators with non-trivial field distributions}, %able to include the effect of a finite resonator size with a non-uniform mode function,
which extends the applicability beyond state-of-the-art work based on \emph{ab-initio} methods~\cite{latini2019cavity}.
We \rev{also note} %also note
that recent theoretical work has analyzed the possibility of obtaining polariton blockade~\cite{kyriienko2020nonlinear} with trions in two-dimensional semiconductors, although without considering any dephasing mechanisms, which we find to be the most important limitation for polariton blockade in realistic systems.
The excitons in a sheet of 2D semiconductor can be approximated as interacting bosons with annihilation operators $\hat{b}_\mathbf{k}$, labeled by the lateral center-of-mass momentum, $\mathbf{k}=(k_x,k_y)$~\cite{usui1960excitations,hanamura1970theory,tassone1999exciton,rochat2000excitonic}. We %shall
approximate the excitonic wave function $\phi(\mathbf{q})$ as the Wannier-Mott type~\cite{cheiwchanchamnangij2012quasiparticle,wang2018colloquium}, having a form corresponding to the hydrogen ground state with exciton Bohr radius $a_{\rm B}$\rev{; this} %. This
approach has previously been demonstrated to agree well with %more
\rev{detailed} numerical calculations~\cite{olsen2016simple}\rev{.} % and \ptk{enables an} %allows for
%analytical treatment of the excitons.
The Hamiltonian of the excitons and \rev{the electromagnetic field} %mode
is split into three parts as $\hat{H} = \hat{H}_{\rm c} + \hat{H}_{\rm x} + \hat{H}_{\rm I}$\ptk{. The first term,}  %. Here,
$\hat{H}_{\rm c} = \hbar\omega_{\rm c} \hat{a}_{\rm c}^\dagger \hat{a}_{\rm c}$\ptk{, represents} the %\ptk{free evolution of the system part of the} % resonant
free evolution of the \ptk{electromagnetic field} %resonator mode
with resonance frequency $\omega_{\rm c}$ and bosonic \rev{annihilation} operator \ptk{$\hat{a}_{\rm c}$~\cite{franke2019quantization}. The second term,} $\hat{H}_{\rm x} = \sum_\mathbf{k} \hbar\omega_\mathbf{k} \hat{b}_\mathbf{k}^\dagger \hat{b}_\mathbf{k} + \hat{W}$\ptk{,} is the exciton Hamiltonian with energies $\hbar\omega_\mathbf{k} = \hbar^2 k^2/(2M) + \hbar\omega_0$, where $M=m_\mathrm{e}+m_\mathrm{h}$ is the total exciton mass ($m_\mathrm{e}$ and $m_\mathrm{h}$ denote the effective electron and hole masses) and $\hbar\omega_0$ is the exciton energy gap, accounting for the exciton binding energy and the bare band gap. The \ptk{operator} $\hat{W}=\sum_{\mathbf{k}\mathbf{k}'\mathbf{q}}\hbar W_{\mathbf{k}\mathbf{k}'\mathbf{q}}\hat{b}_{\mathbf{k}+\mathbf{q}}^\dagger \hat{b}_{\mathbf{k}'-\mathbf{q}}^\dagger \hat{b}_{\mathbf{k}'}\hat{b}_\mathbf{k}$ \ptk{accounts for} %is the
Coulomb-induced exciton--exciton interaction with matrix elements $W_{\mathbf{k}\mathbf{k}'\mathbf{q}}$~\cite{tassone1999exciton,rochat2000excitonic}. \ptk{The third term is the light-matter} %The exciton--resonator
interaction Hamiltonian\ptk{,} $  \hat{H}_{\rm I} = \sum_{\bf k} \hbar (g_{{\bf k}}^* \hat{a}_{\rm c}^\dagger \hat{b}_\mathbf{k} + g_{\mathbf{k}}\hat{a}_{\rm c} \hat{b}_\mathbf{k}^\dagger)$, with coupling strengths~\cite{longpaper}
\begin{align}
\label{eq:g-k}
\hbar g_{\mathbf{k}} = -\frac{e_0}{m_0}\sqrt{\frac{\hbar}{\pi\epsilon_0 \omega_{\rm c} a_{\rm B}^2 S}}\int\dd[2]{\mathbf{r}} e^{-i\mathbf{k}\cdot\mathbf{r}}\tilde{\mathbf{F}}_{\rm c}(\mathbf{r},z_0)\cdot\mathbf{p}_{\rm cv},
\end{align}
where $e_0$ is the elementary charge, $m_0$ is the free electron mass, $S$ is the surface area of the 2D material sheet \ptk{located at $z=z_0$, and $\mathbf{r}=(x,y)$ is the lateral position.} $\tilde{\mathbf{F}}_{\rm c}(x,y,z)$ \ptk{is the field profile of the resonant electric field}\rev{,} % is the regularised electric field profile of the resonator quasi-normal mode, $\tilde{\mathbf{f}}_{\rm c}$ (with the 2D sheet located at $z=z_0$), $\mathbf{r}=(x,y)$ is the lateral position
and $\mathbf{p}_{\rm cv}$ is the momentum Bloch matrix element. %
%
%\ptk{Inside or close to the resonator, the electromagnetic field of interest is described by a % in optical cavities or plasmonic particles are
%quasinormal mode~\cite{}, also known as a resonant state~\cite{}, defined as a %which is calculated as the
%solutions to the sourceless wave equation %(omitting the contribution to the dielectric function from the excitons)
%subject to the radiation condition of free space. It has complex eigenfrequency, $\omega_{\rm c} - \text{i}\gamma_{\rm c}$, where $\omega_{\rm c}$ and $\gamma_{\rm c}$ are the resonance frequency and decay rate, respectively~\cite{kristensen2014modes,franke2019quantization}. The extension of the formalism to positions in the 2D sheet results in an analytical continuation of the QNM $\tilde{\mathbf{f}}_{\rm c}$ onto the real axis~\cite{}, as detailed in Ref.~\cite{longpaper}. } %are calculated from a non-Hermititian eigenvalue problem.
%in the resonator is modeled with vcelectromagnetic field profile pertaining to the is }
%
We also include in the formalism the possibility that the fundamental exciton mode is degenerate, as in transition metal dichalcogenides~\cite{longpaper}.

%The quasi-normal mode $\tilde{\mathbf{f}}_{\rm c}$ is calculated from a non-Hermitian eigenvalue problem (omitting the contribution to the dielectric function from the excitons) and has the complex eigenfrequency, $\omega_{\rm c} - i\gamma_{\rm c}$, where $\omega_{\rm c}$ and $\gamma_{\rm c}$ are the resonance frequency and decay rate, respectively~\cite{kristensen2014modes,franke2019quantization}.

%To ensure convergence of the electric field of the mode far away from the resonator and a well-defined value of the integral in Eq.~\eqref{eq:g-k}, a regularization of the quasi-normal mode, $\tilde{\mathbf{f}}_{\rm c}$ based on the quasi-static Green's function is used to calculate $\tilde{\mathbf{F}}_{\rm c}$.

\begin{figure}
  \centering
  \includegraphics[width=\columnwidth]{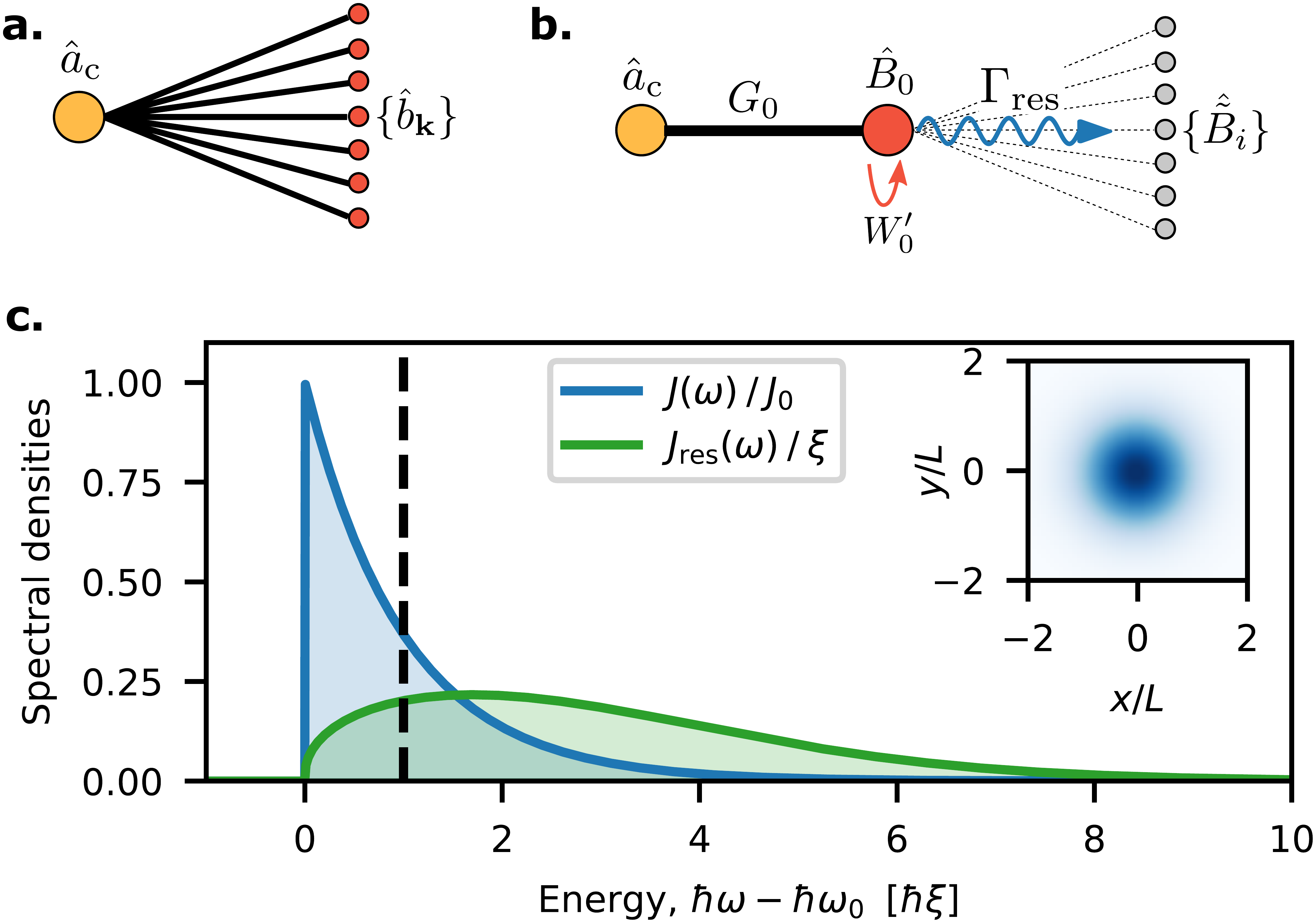}
  \caption{{\bf a.} The localised \rev{resonant field} %resonator mode
  ($\hat{a}_{\rm c}$, orange dot) is coupled to a continuum of exciton modes with momentum $\mathbf{k}$. {\bf b.} Through a linear transformation, the light--matter interaction can be described as a coupling between the \rev{resonant field} %resonator mode
and a single, collective exciton reaction coordinate ($\hat{B}_0$) which, in turn, is coupled to an environment of residual exciton modes ($\hat{\tilde{B}}_i$). %The exciton decay into the residual environment can often be approximated as a Markovian process with rate $\Gamma_{\rm res}$. Furthermore, nonlinear exciton-exciton interactions with rate $W_0'$ can be accounted for within the reaction coordinate.
{\bf c.} Exciton spectral density (blue) and residual spectral density (green) for an \rev{electromagnetic field} %optical mode
with lateral Gaussian confinement corresponding to Eq.~\eqref{eq:gaussian-spec-dens}. The exciton reaction coordinate frequency, $\Omega_0$, is indicated with a dashed black line. The inset shows the Gaussian lateral \rev{field} %mode
profile.
% {\bf b.} Exciton spectral density and residual spectral density (scaled by $10^4$) for a gold nanorod coupled to a $\mathrm{WS_2}$ monolayer sheet located 5 nm from the surface of the nanorod. The inset shows the quasi-normal mode field function in the $\mathrm{WS_2}$ sheet, projected onto the circular polarisation of the Bloch momentum matrix element, $\abs{\tilde{\mathbf{f}}_{\rm c}(\mathbf{r},0)\cdot\mathbf{p}_{\rm cv}}^2.$
}
  \label{fig:spectral-densities}
\end{figure}

%\begin{figure}
%  \centering
%  \includegraphics[width=\columnwidth]{intro-fig.pdf}
%%  \includegraphics[width=\columnwidth]{Fig_1_frontpage.jpg}
%  \caption{{\bf a.} Illustration of optical resonator coupled to a sheet of 2D material. {\bf b.} The localised resonator mode ($\hat{a}$, orange dot) is coupled to a continuum of exciton modes labelled by their in-plane center-of-mass momentum $\mathbf{k}$. {\bf c.} Through a linear transformation, the light--matter interaction can be described as a coupling between the resonator mode and a single, collective exciton reaction coordinate ($\hat{B}_0$), which is in turn coupled to an environment of residual exciton modes ($\hat{\tilde{B}}_i$).
%    %The exciton decay into the residual environment can often be approximated as a Markovian process with rate $\Gamma_{\rm res}$. Furthermore, nonlinear exciton-exciton interactions with rate $W_0'$ can be accounted for within the reaction coordinate.
%    {\bf d.} Exciton--resonator coupling $G_0$, nonlinear exciton-exciton interaction, $W_0'$ and decay rate into residual exciton modes, $\Gamma_{\rm res}$ as a function of the lateral confinement of the optical mode, $L$ for $\mathrm{WS_2}$ coupled to a Gaussian resonator mode with $\epsilon L_z=150\mathrm{\: nm},\; \abs{\mathbf{p}_{\rm cv}\cdot\mathbf{n}}/p_{\rm cv}=0.2$.
%}
%  \label{fig:intro}
%\end{figure}
\ptk{Through Eq.~(\ref{eq:g-k}), the interaction Hamiltonian defines the coupling between the resonant electromagnetic field and each of the free excitons, as illustrated in Fig.~\ref{fig:spectral-densities}\rev{a.} %b.
\rev{To simplify this model, we can} %We
perform a change of basis by defining the} %define a
%The asymmetric coupling %By definition, the reaction coordinate is the sum of free excitons weighted by their interaction with the resonant electric field.
%Defining the
collective exciton reaction coordinate with bosonic annihilation operator $\hat{B}_0 = G_0^{-1}\sum_\mathbf{k} g_{\mathbf{k}}^* \hat{b}_\mathbf{k}$, where $G_0=\big(\sum_\mathbf{k} \abs*{g_{\mathbf{k}}}^2\big)^{1/2}$. \ptk{In this way we can} %, we can
write the light--matter interaction compactly as $\hat{H}_{\rm I} = \hbar G_0(\hat{B}_0^\dagger \hat{a}_{\rm c} + \hat{B}_0 \hat{a}_{\rm c}^\dagger)$. The collective coupling strength is related to the \ptk{electric field} %resonator mode
profile as
\begin{align}
\label{eq:G0}
G_0^2 = \frac{e_0^2}{\pi\hbar\epsilon_0 m_0^2\omega_{\rm c} a_\mathrm{B}^2}\int\dd[2]{\mathbf{r}\abs*{\tilde{\mathbf{F}}_{\rm c}(\mathbf{r},z_0)\cdot\mathbf{p}_{\rm cv}}^2}.
\end{align}
Importantly, this coupling strength depends on the field intensity integrated over the entire 2D-material surface. \ptk{Therefore, in contrast to what one would expect in the dipole approximation~\cite{stobbe2012spontaneous}, tightening the \rev{electromagnetic} %optical
confinement in the lateral direction does not lead to an increased coupling strength, as seen by the horizontal black solid line in Fig.~\ref{fig:Fig_1_frontpage}, where separability of the \rev{field} %mode
in the lateral and out-of-plane coordinates was assumed. Instead, } %, as one would expect in the dipole approximation~\cite{stobbe2012spontaneous}. Rather,
the overall field strength in the \rev{plane of the 2D-material} %plane
can be quantified through the out-of-plane confinement length scale, $L_z=(\int\dd[2]{\mathbf{r}}\abs*{\tilde{\mathbf{F}}_{\rm c}(\mathbf{r},z_0)}^2)^{-1}$. \ptk{Indeed, $L_z$ defines an} % This length scale, $L_z$, is then related to the
upper bound of the coupling strength through the inequality $G_0^2 \leq  e_0^2\abs{\mathbf{p}_{\rm cv}}^2/[\hbar\pi\epsilon_0 m_0^2 \omega_{\rm c} a_{\rm B}^2 L_z]$, which becomes an equality when the polarisation of $\tilde{\mathbf{F}}_{\rm c}(\mathbf{r},z_0)$ is aligned with $\mathbf{p}_{\rm cv}$. \ptk{In order to implement the change of basis and benefit from} %Despite
the simpler form of $\hat{H}_{\rm I}$, we must reformulate the noninteracting exciton Hamiltonian $\hat{H}_{\rm x,0}:=\sum_\mathbf{k} \hbar\omega_\mathbf{k} \hat{b}_\mathbf{k}^\dagger \hat{b}_\mathbf{k}$
by defining a new set of exciton modes with bosonic operators $\hat{\tilde{B}}_{i}$ and frequencies $\tilde{\Omega}_i$ ($i>0$), such that it can be written as
\begin{align*}
\hat{H}_{\mathrm{x},0}
  = \hbar\Omega_0 \hat{B}_{0}^\dagger \hat{B}_0 + \sum_{i>0}\qty[\hbar\tilde{\Omega}_i\hat{\tilde{B}}_i^\dagger \hat{\tilde{B}}_i  + ( \hbar\tilde{\lambda}_i \hat{B}_0^\dagger \hat{\tilde{B}}_i + \mathrm{H.c.})],
\end{align*}
\rev{where} %with the reaction coordinate frequency
$\Omega_0=\sum_\mathbf{k} \abs{g_{\mathbf{k}}}^2\omega_\mathbf{k}/G_0^2$ \rev{is the reaction coordinate frequency,} % \ptk{,}
and where the \rev{operators fulfill canonical} commutation relations\rev{,} % are canonical,
$[\hat{\tilde{B}}_i,\hat{\tilde{B}}_j^\dagger]=\delta_{ij}$ and $[\hat{B}_0,\hat{\tilde{B}}_i^\dagger]=0$. Such a transformation is described within the theory of reaction coordinate mappings~\cite{leggett1984quantum,garg1985effect,hughes2009effective,hughes2009effective2,iles2016energy}.
This form of \ptk{$\hat{H}_{\mathrm{x}}$} %$\hat{H}_{\mathrm{x},0}$
describes the interaction of the exciton reaction coordinate $\hat{B}_0$ with an environment of residual exciton modes $\hat{\tilde{B}}_i$ through the coupling coefficients $\tilde{\lambda}_i$\ptk{, as} %. This representation is
depicted in Fig.~\ref{fig:spectral-densities}b. %Unlike previous work~\cite{verger2006polariton, tserkezis2020applicability}, \ptk{the present} approach precisely identifies and accounts for these residual exciton modes.
By identifying the residual exciton modes, we can calculate the decay rate into the residual environment in the Markovian limit \rev{and} %, and to
assess the conditions under which the excitons can be described in terms of a single \ptk{reaction coordinate.} %isolated mode.
The properties of the residual environment are contained in the residual spectral density, $J_\mathrm{res}(\omega) = \sum_{i>0} \abs*{\tilde{\lambda}_i}^2\delta(\omega-\tilde{\Omega}_i)$, which is
%which \ptk{is} %can be
related to the exciton spectral density $J(\omega)=\sum_\mathbf{k}\abs{g_{\mathbf{k}}}^2\delta(\omega-\omega_\mathbf{k})$ as~\cite{martinazzo2011communication,woods2014mappings}
\begin{align}
  J_{\rm res}(\omega) = \frac{G_0^2J(\omega)}{\Phi^2(\omega) + \pi^2J^2(\omega)},
\end{align}
where $\Phi(\omega)= \lim_{\ell\rightarrow 0^+}\int_{\omega_0}^\infty\dd{z}J(z)\frac{\omega-z}{(\omega-z)^2 + \ell^2}$.

To quantify the impact of the optical confinement geometry, we now assume that the \rev{electromagnetic field profile} %resonator mode
is separable in the lateral and out-of-plane coordinates %,
and that the \rev{field} %mode
is uniformily polarised with polarisation vector $\mathbf{n}$, such that $\tilde{\mathbf{F}}_{\rm c}(\mathbf{r},z) = \mathbf{n} F_z(z) F_\|(\mathbf{r})$. \rev{In this case,} %Here,
the out-of-plane confinement length takes the intuitive form $L_z = \epsilon_{\rm eff}\int \dd{z}\abs{F_z(z)}^2/\abs{F_{z}(z_0)}^2$, where $\epsilon_{\rm eff}$ is an effective dielectric constant \rev{accounting} %that accounts
for the dielectric \rev{environment. In this way,} %surroundings. Thus,
$L_{z}$ can %in this case
be formally separated from the lateral \rev{field distribution.} % due to the separability of $\tilde{\mathbf{F}}_{\rm c}$.
Furthermore, when taking the in-plane field distribution to be Gaussian with confinement length $L$; $F_{\|}(\mathbf{r}) = e^{-r^2/(2L^2)}/\sqrt{\pi L^2}$, we can evaluate the spectral densities analytically as~\cite{longpaper}
\begin{align}
\label{eq:gaussian-spec-dens}
\begin{split}
 J(\omega) &= \Theta(\omega-\omega_0)J_0 e^{-(\omega-\omega_0)/\xi}, \\
J_\mathrm{res}(\omega) &= \Theta(\omega-\omega_0)\xi e^{(\omega-\omega_0)/\xi}\{\mathrm{Ei}^2[(\omega-\omega_0)/\xi] + \pi^2\}^{-1},
\end{split}
\end{align}
\comment{Emil: skal vi ikke droppe ``+''-tegnet i exponenten i den anden ligning. Desuden: Er der styr på fortegnet i Ei? ... var der ikke noget med, at det skal være omvendt?}
where $\Theta$ is the Heaviside function, $\xi = \hbar /(2 ML^2)$ is a cutoff frequency, $J_0 = G_0^2/\xi$ determines the overall magnitude of the spectral density and \ptk{$\mathrm{Ei}$ denotes} %$\mathrm{Ei}(x)=\int_{-\infty}^x\dd{z}\exp(z)/z$ is
the exponential integral\rev{; these spectral densities are shown in} % In
Fig.~\ref{fig:spectral-densities}c\rev{.} %,  the spectral densities \ptk{of} Eq.~\eqref{eq:gaussian-spec-dens} are shown.
The overall scaling of the residual spectral density is determined by $\xi\propto L^{-2}$\ptk{, which shows that} %; this already shows that
the residual excitons are most important when the lateral confinement length is small.

\begin{figure}
  \centering
\includegraphics[width=\columnwidth]{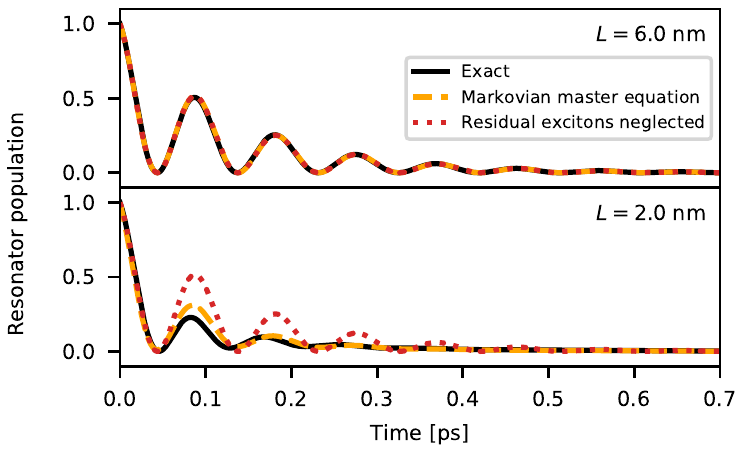}
  \caption{{\bf a.} Exact time evolution of resonator \ptk{excitation} %photon
  number, $\ev*{\hat{a}_\mathrm{c}^\dagger \hat{a}_\mathrm{c}}$ (black solid lines), compared with the Markovian theory (orange dashed lines) and with the residual excitons ignored (red dotted lines) for a separable \rev{field profile} %mode
with Gaussian in-plane \rev{distributions and} %mode function at
different lateral confinement lengths, $L$. The 2D material was taken to be $\mathrm{WS_2}$, and $L_z=150\mathrm{\; nm},\; \omega_{\rm c}=\Omega_0,\; \hbar\gamma_{\rm c}=5\mathrm{\;meV}$, \ptk{and} $\abs{\mathbf{n}\cdot\mathbf{p}_{\rm cv}}/p_{\rm cv}=0.5$. The light-matter coupling with these parameters is $\hbar G_0=22\mathrm{\: meV}$.
  }
  \label{fig:residual-decoupling}
\end{figure}
\ptk{To assess the limits of the model based on a single reaction coordinate, we investigate the} %An important task is to determine the
conditions for neglecting the residual excitons. % environment.
To this end, we study the linear response limit, where the nonlinear interactions $\hat{W}$ can be ignored \ptk{and} %. \ptk{We} %Here, we
consider the time evolution of a single resonator excitation. The excitation number $\ev*{\hat{a}_\mathrm{c}^\dagger(t) \hat{a}_\mathrm{c}(t)}$ can %then
be calculated exactly through the equation for the resonator amplitude $\phi_{\rm c}(t)$~\cite{vats2002theory},
\begin{align}
\dv{\phi_{\rm c}(t)}{t} = -\int_0^tK(t-t')\phi_{\rm c}(t')\dd{t'} - \gamma_{\rm c}\phi_{\rm c}(t),
\end{align}
where $K(\tau) = \Theta(\tau)\int\dd{\omega}J(\omega) e^{-i(\omega-\omega_{\rm c})\tau}$ is a memory kernel, which fully accounts for interactions with the excitons, and $\gamma_\text{c}$ is the decay rate of the electromagnetic field. The excitation number of the resonator is then given by $\ev*{\hat{a}_\mathrm{c}^\dagger(t) \hat{a}_\mathrm{c}(t)} = \abs{\phi_{\rm c}(t)}^2$. This time evolution is shown in Fig.~\ref{fig:residual-decoupling} (black solid). As an alternative approach, we have derived a master equation for the reduced density operator of the \rev{resonant field} %resonator mode
and exciton reaction coordinate, \rev{in which} %where
the effect of the residual excitons is approximated by a Markovian decay with rate $\Gamma_\mathrm{res} = 2\pi J_\mathrm{res}(\omega_+)$, \ptk{where} $\omega_+=[\omega_{\rm c}+\Omega_0 + \sqrt{4G_0^2+(\omega_{\rm c}-\Omega_0)^2}]/2$ \ptk{is} the frequency of the upper polariton~\cite{longpaper}. \rev{The} %This
residual decay rate is shown in Fig.~\ref{fig:Fig_1_frontpage}b
for $\omega_{\rm c}=\Omega_0$\rev{, and the time} %. The time
evolution generated by this master equation is shown in Fig.~\ref{fig:residual-decoupling} (orange dashed) along with the result obtained when ignoring the residual excitons entirely (red dotted). %Fig.~\ref{fig:residual-decoupling}b shows the integrated mean square error between the exact result and the Markovian master equation, as well as the case where the residual modes are neglected.
This shows that the residual exciton environment starts to play a role for $L$ below $\sim 4 \mathrm{\:nm}$, and that it is well approximated by the Markovian theory~\cite{longpaper}.
%These calculations show that above a certain lateral length scale, here above 3 nm, the residual excitons can be ignored as a very good approximation. It is also demonstrated that the Markovian approximation for the interaction with the residual excitons provides a useful improvement of the reaction coordinate master equation in the parameter regime where influence of the residual environment starts to become important.
%For the numerically calculated quasi-normal mode from Fig.~\ref{fig:spectral-densities}b, coupled to $\mathrm{WS_2}$, we find $\hbar\Gamma_\mathrm{res}=170\mathrm{\; \mu eV}$ and a relative error of $4.91\times 10^{-4}$ in the Markovian master equation, and $4.96\times 10^{-4}$ when the residual excitons are ignored entirely.
% The coupling strength is $\hbar G_0 = 36.4\mathrm{\; meV}$, the cavity decay rate is $\hbar \gamma_{\rm c} = 228\mathrm{\; meV}$, and the detuning is $\delta_{cx}=-7.2 \mathrm{\; meV}$.

%\paragraph{Nonlinear exciton-exciton interactions and polariton blockade \textemdash}
\begin{figure}
  \centering
  \includegraphics[width=\columnwidth]{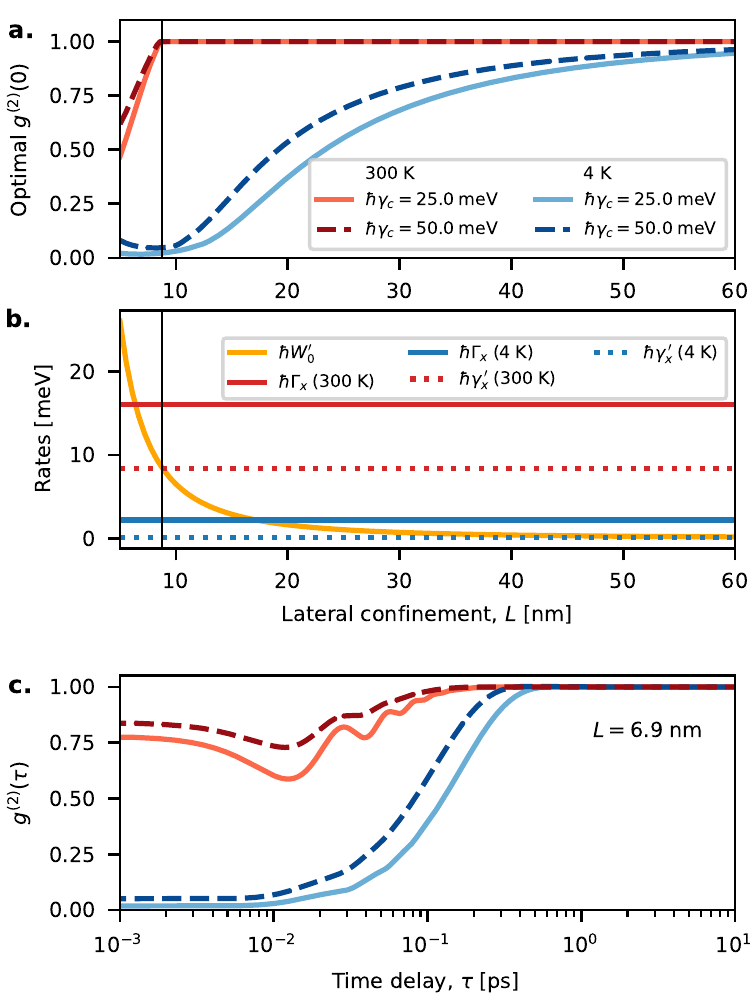}
  \caption{{\bf a.} Optimal antibunching versus lateral confinement, $L$, for monolayer $\mathrm{WS_2}$ coupled to a \rev{resonant field with} Gaussian in-plane \rev{distribution} %resonator mode
at temperatures of 4 K and 300 K and resonator linewidths $\hbar\gamma_{\rm c}$ of $25\:\mathrm{meV}$ and $\:50\mathrm{\:meV}$. Parameters: $L_z=50\mathrm{\: nm},\:F=1.5\mathrm{\:meV},\; \abs{\mathbf{p}_{\rm cv}\cdot\mathbf{n}}/p_{\rm cv}=0.75$, corresponding to $\hbar G_0 = 57.5\mathrm{\: meV}$. {\bf b.} Nonlinear interaction strength $W_0'$ versus $L$, compared to exciton linewidth, $\Gamma_{\rm x}=\gamma_{\rm x}+\gamma_{\rm x}^{\:\prime}$, and exciton dephasing, $\gamma_{\rm x}^{\:\prime}$. The thin vertical lines in panels a and b mark the value of $L$ where $g^{(2)}(0)$ drops below 1 for $T=300\mathrm{\:K}$.  {\bf c.} Delay-time dependence of steady-state second-order correlation function for optimized parameters as in panel a, at a lateral confinement length of $L=6.9\;\mathrm{nm}$. Plot signatures correspond to those in panel a.}
  \label{fig:nonlinear}
\end{figure}

We now proceed to study the impact of the nonlinear exciton-exciton interactions and the prospects for reaching \ptk{the} polariton blockade \ptk{regime}. To do so, we assume that the residual exciton modes, $\hat{\tilde{B}}_{i}$, are weakly populated, such that the exciton--exciton interaction, $\hat{W}$, is only significant within the reaction coordinate. This assumption is reasonable in the limit where the exciton reaction coordinate is decoupled from the residual exciton modes and \rev{therefore} is %thus
the only exciton mode with appreciable population. The \rev{interaction is} %term $\hat{W}$ is
thus approximated as \rev{$\hat{W}\approx\hat{W}_0:=\hbar W_0' \hat{B}_0^\dagger \hat{B}_0^\dagger \hat{B}_0 \hat{B}_0$. Assuming $L\gg a_{\rm B}$,} %\ptk{, where} %. Here
%$W_0'$ is the interaction strength within the reaction coordinate. In the regime where $L\gg a_{\rm B}$,
we can neglect the momentum-dependence of the matrix element $W_{\mathbf{k}\mathbf{k}'\mathbf{q}}$~\cite{tassone1999exciton}, such that \rev{ the interaction strength within the reaction coordinate becomes} $W_0' \simeq SW_{000}\eta_\mathbf{n}[\int\dd[2]{\mathbf{r}} \abs*{\tilde{\mathbf{F}}_{\rm c}(\mathbf{r},z_0)\cdot\mathbf{p}_{\rm cv}}^2]^{-2}\int\dd[2]{\mathbf{r}} \abs*{\tilde{\mathbf{F}}_{\rm c}(\mathbf{r},z_0) \cdot\mathbf{p}_{\rm cv}}^4$ \cite{verger2006polariton,longpaper}. This quantity determines the nonlinear energy shift of the exciton reaction coordinate. Here, the polarization-dependent prefactor $\eta_\mathbf{n}$ takes values between 1/2 (reached in the limit of a linearly polarized resonator mode) and 1 (reached in the limit of a circularly polarized resonator mode)~\cite{longpaper}. In the calculations presented here, we take $\eta_\mathbf{n}=1$, corresponding to a circularly polarized mode.

The zero-momentum interaction matrix element has previously been found to be \ptk{well} approximated from the binding energy ($E_{\rm b}$) and Bohr radius as $\hbar W_{000} \simeq \alpha E_{\rm b} a_{\rm B}^2/S$\rev{. Using $\alpha = 2.07$ for $\mathrm{WS_2}$~\cite{shahnazaryan2017exciton}, this leads to} %, where for $\mathrm{WS_2}$ \rev{with $\alpha = 2.07$~\cite{shahnazaryan2017exciton}, leading to
$\hbar S W_{000} \simeq 2.04 \mathrm{\; eV nm^2}$. For the Gaussian \rev{field} %resonator mode introduced above,$
we find $\hbar W_0'= \alpha E_{\rm b} a_{\rm B}^2/(2\pi L^2)$, which scales as the inverse mode area. \rev{In} %We note that, in
contrast to the case of a laterally nanostructured 2D-material~\cite{wang2017quantum,ryou2018strong}, the nonlinear interaction strength here is entirely determined by the lateral confinement \rev{of the resonant field} %length of the resonator mode
and the exciton Coulomb interaction strength. \rev{Within the master equation formalism, we can} %   Furthermore, we
account for non-radiative exciton decay and dephasing due to phonon interactions~\cite{moody2015intrinsic} through %. Within the master equation formalism, we describe these effects
a non-radiative exciton decay rate $\gamma_{\rm x}$, and dephasing rate $\gamma_{\rm x}^{\,\prime}$, which both increase with temperature~\cite{selig2016excitonic,longpaper}.
%   In monolayer transition-metal dichalcogenides, the band-gap is located at the $K/K'$-point~\cite{wang2018colloquium}. Radiative decay and interactions with phonons at the $\Lambda$-point leads to a temperature-dependent decay into optically inaccessible exciton states with a rate $\gamma_{\rm x}$; furthermore, thermal phonons at the $\Gamma$-point generate intra-valley scattering that leads to exciton decoherence with a rate $\gamma_{\rm x}'$~\cite{selig2016excitonic}. We account for these non-radiative phonon-induced effects as Markovian processes in the master equation formalism~\cite{longpaper}.
%One of the most important signatures of nonlinear interactions in this system polariton blockade, i.e. the inhibition of multi-polariton excitation.

\rev{To investigate polariton blockade and single-photon non-linearities, we consider the} %the The signature of polariton blockade, and thus single-photon nonlinearity, is the
second-order correlation function % of the resonator mode,
$g^{(2)}(\tau)=\ev*{\hat{a}_\mathrm{c}^\dagger\hat{a}_\mathrm{c}^\dagger(\tau) \hat{a}_\mathrm{c}(\tau)\hat{a}_\mathrm{c}}/\ev*{\hat{a}_\mathrm{c}^\dagger\hat{a}_\mathrm{c}}^2$~\cite{brown1956correlation}\rev{,} evaluated with respect to the steady-state density operator\rev{.} \ptk{Zero-delay values of} %Here, values
$g^{(2)}(0)<1$ \ptk{are} a signature of polariton blockade, and $g^{(2)}(0)$ reaches 0 in the case of perfect blockade~\cite{gardiner2004quantum,verger2006polariton}. It has previously been shown that in the absence of dephasing, near-zero values of $g^{(2)}(0)$ can be reached, even in the limit of a small nonlinear interaction, $W_0'\ll \gamma_{\rm x}$~\cite{liew2010single,ferretti2013optimal}. This is an unconventional polariton blockade effect, which arises due to destructive interference between the excitation paths leading to excitation of multiple polaritons~\cite{bamba2011origin}. We note that the original theoretical descriptions of unconventional polariton blockade~\cite{liew2010single,bamba2011origin} found the effect in a system of two coupled electromagnetic resonators, with at least one of them being nonlinear. In contrast, the two coupled modes in the present situation are constituted by the resonator and the exciton reaction coordinate, where only the latter is nonlinear. We use the Markovian master equation to evaluate the minimal value of $g^{(2)}(0)$ that can be reached for a given lateral confinement length $L$, of the resonator with a Gaussian electromagnetic \rev{field} profile, driven by a continuous-wave laser field with frequency $\omega_\mathrm{d}$ and amplitude $F$. This driving is described by adding the term $F e^{+i\omega_\mathrm{d} t}\hat{a}_\mathrm{c}+F^* e^{-i\omega_\mathrm{d} t}\hat{a}_\mathrm{c}^\dagger$ to $\hat{H}_{\rm c}$. The nonlinear interaction is accounted for within the exciton reaction coordinate through the Hamiltonian contribution $\hat{W}_0$.
We then numerically minimize $g^{(2)}(0)$ with respect to $\omega_\mathrm{d}$ and $\omega_\mathrm{c}$ for each parameter setting.

\rev{Figure~\ref{fig:nonlinear}a %~In Fig.~\ref{fig:nonlinear}a, \rev{we
shows the minimum attainable} %the minimized
$g^{(2)}(0)$ %is shown
for $\mathrm{WS_2}$ as a function of $L$ at cryogenic temperature (blue colors) and room temperature (red colors)\ptk{, and for} %with
different \rev{experimentally relevant} resonator linewidths $\gamma_{\rm c}$. %, with experimentally relevant values.
In Fig.~\ref{fig:nonlinear}b, we \rev{show} %plot
the nonlinear shift $W_0'$ as a function of $L$ and show the total intrinsic exciton linewidth $\Gamma_{\rm x}=\gamma_{\rm x}+\gamma_{\rm x}^{\,\prime}$, for the two temperature settings, along with the dephasing contribution to the linewidth, $\gamma_{\rm x}'$. From these calculations, we observe the important result that reductions in $g^{(2)}(0)$ below the classical limit $g^{(2)}(0)<1$ \ptk{become} feasible when $W_0'$ overcomes the polariton dephasing $\gamma_{\rm x}^{\:\prime}$, whereas it does not need to overcome the total linewidth. % \comment{der er jo også det radiative bidrag til $\gamma_X$, som ikke er medregnet her (ændrer ikke på sandheden af udsagnet)}.
% \comment{Er der ikke noget galt med denne beskrivelse (ser f.eks. ud til at være $L<80$ (og ikke $L>20$) for T=4K:}
\rev{This follows from} %In Fig.~\ref{fig:nonlinear}, this is observed in
the difference between the behaviour at the two temperatures: For $T=4\mathrm{\:K}$, $W_0'$ is above $\gamma_{\rm x}^{\:\prime}$ for $L\lesssim 75\mathrm{\: nm}$ and $g^{(2)}(0)$ increases smoothly from near-zero values towards 1. At $T=300\mathrm{\: K}$, $W_0'$ exceeds $\gamma_{\rm x}^{\:\prime}$ for $L\lesssim 9\mathrm{\:nm}$, at which point the optimal $g^{(2)}(0)$ starts to decrease from unity abruptly, due to the steep increase in $W_0'$ as $L$ is decreased (indicated with thin vertical lines in Fig.~\ref{fig:nonlinear}a-b). \rev{In this way, we identify excitonic dephasing as the main challenge \ptk{for %that needs to be overcome in order to
reaching} the blockade regime\ptk{, and we} %We
interpret this limitation as originating from a decoherence-induced suppression of the destructive interference between two-polariton exciton paths, which is responsible for polariton blockade.} In Fig.~\ref{fig:nonlinear}c, we show the delay-time dependence of $g^{(2)}(\tau)$\rev{, as calculated} % (calculated
from the master equation using the quantum regression theorem~\cite{breuer2002theory}, which transitions from its minimal value at $\tau=0$ towards unity on a time scale on the order of $1\:\mathrm{ps}$. \rev{The antibunching appearing on this picosecond time scale should be observable with established ultrafast detection techniques~\cite{mcalister1997ultrafast,assmann2009higher,assmann2010measuring}. Due to a significant detuning on the order of 100-300 meV between the \rev{resonant field} % resonator mode
and the exciton in the two-frequency optimized configuration, the signal does not exhibit Rabi oscillations as reported in previous studies on unconventional polariton blockade~\cite{liew2010single,bamba2011origin,ferretti2013optimal}. This significant detuning follows from the fact that our optimization procedure does not impose constraints on the frequencies $\omega_{\rm c}$ and $\omega_{\rm d}$, whereas previous studies have fixed one of these detunings in the optimization procedure.
}

%The significant detuning present in the optimal configuration was not reported in previous studies on unconventional polariton blockade~\cite{liew2010single,bamba2011origin,ferretti2013optimal}.

%\paragraph{Conclusion\textemdash}
We have presented a microscopic theory of light-matter interaction in electromagnetic resonators coupled to pristine sheets of 2D semiconductor, whereby we unambiguously and analytically identify all relevant dynamical parameters in terms of fundamental material parameters.
% which provides analytical relations between the relevant dynamical parameters and the fundamental material parameters.
\rev{We find} that %Thereby, we find that
rather than the mode volume $\sim L^2 L_z$, it is the out-of plane confinement length $L_z$ that controls the light-matter interaction, while the lateral length $L$ determines both the nonlinear interaction and coupling to residual modes. Using this theory, we have identified a significant operational window of lateral length scales, where the nonlinear exciton response is large enough to \ptk{enable a pronounced} %allow for
polariton blockade \ptk{while the} %and
coupling to residual modes can be neglected. \rev{Our calculations \ptk{show} that lateral dimensions of the order \ptk{of} $40\mathrm{\:nm}$ or smaller are required to see significant antibunching, $g^{(2)}(0)<1/2$, at cryogenic temperatures, whereas smaller dimensions of around $10\mathrm{\:nm}$ are required at room temperature. Such dimensions are well within the limits of contemporary nanofabrication~\cite{jessen2019lithographic} using either plasmonic resonators~\cite{duan2012nanoplasmonics,benz2016single} or optical cavities with extreme dielectric confinement of light \cite{Hueaat2355, albrechtsen2021nanometer}.} %We believe that these results will \ptk{enable the} %allow for the design \ptk{and modeling} of new experiments demonstrating single-photon nonlinearities with transition-metal dichalcogenides.

%The nonlinear exciton-exciton interaction generated by $\hat{W}_0$ can lead to polariton blockade if the nonlinearity becomes comparable to the polariton linewidth, such that the appearance of an exciton in the reaction coordinate shifts the exciton transition energy sufficiently to reduce further excitations~\cite{verger2006polariton,ferretti2012single,delteil2019towards}.

%Parametric oscillation:~\cite{wouters2007parametric}.

\begin{acknowledgments}
\paragraph{Acknowledgments\textemdash}
The authors thank Peder Meisner Lyngby for valuable discussions. This work was supported by the Danish National Research Foundation through NanoPhoton - Center for Nanophotonics, grant number DNRF147 and Center for Nanostructured Graphene, grant number DNRF103. NS acknowledges support from the Villum Foundation through grant number 00028233. \ptk{MW and NS acknowledge support from the Independent Research Fund Denmark - Natural Sciences (project no. 0135-00403B)}. EVD acknowledges support from Independent Research Fund Denmark through an International Postdoc fellowship (grant no. 0164-00014B).
\end{acknowledgments}

\end{document}